\documentclass{article}
\usepackage[utf8]{inputenc}
\usepackage{amsmath}
\usepackage{graphicx}
\usepackage{enumitem}
\usepackage{mathrsfs}
\usepackage{amssymb}

\usepackage{xcolor}

\title{\bf The Radiation Gauge: When is it Valid?}
\author{J. Zhu, C.J. Ryu, D.Y. Na, W.C. Chew
}

\renewcommand{\v}[1]{{{\bf #1}}}

\def\ED{\end{document}}

\def\be{\begin{equation} }
\def\ee{\end{equation} }

\def\t{\text}

\newcount\timehh\newcount\timemm
\timehh=\time \divide\timehh by 60 \timemm=\time \count255=\timehh
\multiply\count255 by -60 \advance\timemm by \count255
\def\draftheader{\slshape\today\ at
\ifnum\timehh<10 0\fi\number\timehh\,:\,\ifnum\timemm<10 0\fi\number\timemm}%

\renewcommand{\v}[1]{{{\bf #1}}}

\def\curl{\nabla\times}

\def\div{\nabla\cdot}

\def\grad{\nabla}

\def\prw{(\v r, \omega)}
\def\prpw{(\v r', \omega)}

\def\rec2pi{\frac1{2\pi}}

\def\eeqa{\end{eqnarray}}
\renewcommand{\v}[1]{{{\bf #1}}}

\def\be{\begin{equation} }
\def\ee{\end{equation} }

\def\RHS{right-hand side}

\def\Meq{Maxwell's equations}

\def\today{\ifcase\month\or January\or February\or March\or April\or
May\or June\or July\or August\or September\or October\or November\or
December \fi\space\number\day, \number\year}

\def\~{\tilde}
\def\^{\hat}

\def\Div{\nabla\cdot}

\begin{document}
\maketitle

{\let\thefootnote\relax\footnotetext{\textsl{Printed on }\draftheader.}}

\section{Introduction}

Gauge is an important concept in electromagnetic theory \cite{jackson1999classical}.  In fact, it was first used by Maxwell in his original work, because he used the vector and scalar potential formulation \cite{yaghjian2014reflections}.
Recently, the vector-scalar potential formulation has been resuscited in computational electromagneics because it has a number of advantages \cite{chew2014vector,li2009finite,li2015vectorial,li2016finite,li2016loop,liu_qin2018potential,roth2020radiation}.  First, it is immune to low-frequency breakdown.  Namely, when such a method is applied, the system equation is solvable even when the frequency is low or the mesh density is high.  Therefore, this can help with the convergence and robustness of the solution when iterative methods are applied.  Also, the vector-scalar potential formulation merges well with the quantum formulation of electromagnetic theory
\cite{CHEW2016quantum}.

In this paper, we shall show that the vector-scalar potential ($\v A$-$\Phi$) formulation, for many problems, can be further simplified by ignoring the scalar potential contribution and setting it to zero.  While this is not possible for all problems, it is possible for a large class of problems.
  When the scalar potential is ignored by setting it to zero, the Lorenz gauge is the same as the Coulomb gauge.  (It also removes some unpleasant physics of the simple Coulomb gauge where the scalar potential is not zero, resulting in a formulation that is not relativistically invariant as discussed in \cite{jackson1999classical}.)
Such a gauge is sometimes called the radiation gauge, the $\Phi=0$ gauge, or the Weyl gauge (the radiation gauge is a gauge often used in general relativity) \cite{lancaster2014quantum}.

Because of the absence of low-frequency breakdown, $\v A$-$\Phi$ formulation can be used to compute the solutions of multi-scale structures more easily
 \cite{li2016finite,vico2016decoupled,liu2018potential,roth2018development,roth2020stability}.  Moreover, in many applications such as quantum electromagnetics \cite{CHEW2016quantum}, the vector and scalar  potentials, $\v A$ and $\Phi$ are more fundamental than the electric and magnetic fields, $\v E$ and $\v H$.  

%

\section{Free Space Case}

If the medium is free space which is homogeneous, linear, and dispersionless, (assuming $e^{-i\omega t}$ time convention) \Meq\ in the frequency domain become
\begin{align}
\nabla\times{\bf E}=&i\omega\mu_0{\bf H}\label{eq 9}\\
\nabla\times{\bf H}=&-i\omega\varepsilon_0{\bf E}+\v J_{\t{imp}} \\
\nabla\cdot\mu_0{\bf H}=&0\\
\nabla\cdot{\varepsilon_0}{\bf E}=&\varrho_{\t{imp}}\label{eq 12}
\end{align}
where $\v J_{\t{imp}}$ and $\varrho_{\t{imp}}$ are impressed sources: They are immutable in the presence of their environment, and generate the incident field in many scattering problems.\footnote{This trick is used to decouple two systems that are otherwise coupled together. This is similar to the use of alternating direction method to decouple two dynamic systems \cite{boyd2011distributed}.}
The above equations are solvable by using the vector-scalar potential approach by letting
\begin{align}\label{pezg:eq3}
\v B&=\mu_0\v H=\curl \v A,\qquad \\\v E&=i\omega \v A-\grad \Phi
\label{pezg:eq3b}
\end{align}
In this case, as shown in many textbooks, e.g. \cite{kong2008ewt,jackson1999classical}, one way to solve the above equations uniquely is by invoking the Lorenz gauge.

\subsection{Lorenz Gauge}

The Lorenz gauge is the preferred gauge because it treats space and time on equal footing:  Hence, it is relativistically invariant \cite{jackson1999classical}.
Starting with the four Maxwell's equations,
with the use of  Lorenz gauge where
\begin{align}\label{pezg:eq0}
\Div \v A=i\omega \mu_0\varepsilon_0 \Phi
\end{align}
we arrive at two second order equations \cite{chew2014vector},
\begin{align}\label{pezg:eq1a}
    \nabla\cdot\nabla\Phi\prw&+\omega^2\mu_0\varepsilon_0 \Phi\prw=-\varrho_\t{imp}\prw/\varepsilon_0\\
    -\nabla\times\nabla\times{\bf{A}\prw}&+\nabla(\nabla\cdot{\bf{A}\prw})+\omega^2\mu_0\varepsilon_0 {\bf{A}\prw}
    =-\mu_0\bf{J}_\t{imp}\prw,
\label{pezg:eq2a}
\end{align}
where $\v A\prw$ and $\Phi\prw$ are the vector and scalar potentials, respectively, in the frequency domain.
By using $-\nabla\times\nabla\times{\bf{A}\prw}+\nabla(\nabla\cdot{\bf{A}\prw})=\div\grad \v A\prw$, the above can be rewritten as
\begin{equation}
\div\grad \v A\prw+\omega^2\mu_0\varepsilon_0 {\bf{A}\prw}=-\mu_0\bf{J}_\t{imp}\prw
\end{equation}
In
\eqref{pezg:eq2a},
when converted back to the time domain where $\omega^2\rightarrow -\partial_t^2$, the above equations put space and time on the same footing.  Hence, they are relativistically invariant \cite{jackson1999classical}.  Furthermore, it is a common practice to regard the quadruplet, $\left\{\v A,\Phi\right\}$ to be a four-vector in special relavitiy \cite{jackson1999classical}.
%
  In addition, in \eqref{pezg:eq1a} and \eqref{pezg:eq2a},
  \begin{align}
    \Div \v J_\t{imp}\prw=i\omega \varrho_\t{imp}\prw
  \end{align}
    is the current continuity equation
    which is a statement of charge conservation.   In other words, the current $\v J_\t{imp}$ and the charge $\varrho_\t{imp}$ are not independent of each other.
  And, \eqref{pezg:eq1a} and \eqref{pezg:eq2a} are not independent of each other:  \eqref{pezg:eq1a} can be derived from \eqref{pezg:eq2a} by taking its divergence, and using the Lorenz gauge \eqref{pezg:eq0}.  When $\omega\ne 0$, it suffices to solve only \eqref{pezg:eq2a} and obtain \eqref{pezg:eq1a} by invoking the Lorenz gauge and the current continuity equation.   But $\Phi$ cannot be retrieved from $\v A$ by using \eqref{pezg:eq0} when $\omega=0$, or when $\omega\rightarrow 0$.  Hence, both equations \eqref{pezg:eq1a} and \eqref{pezg:eq2a} have to be solved together when $\omega=0$.  This is similar to the notion that all four \Meq\ have to be solved when $\omega=0$ \cite{ece604lectures}.

  The Laplacian operator $\div\grad=\nabla^2$ is a negative definite operator, and hence, it has no low-frequency breakdown, implying that the above equations have solutions no matter how small $\omega$ is or how low the frequency is.\footnote{The Laplacian operator does have an issue with fine-mesh ill-conditioning, which can be remedied with the multi-grid method \cite{hiptmair1998multigrid,axelson1984finite}.}  To derive the above two equations \eqref{pezg:eq1a} and \eqref{pezg:eq2a}, all four of Maxwell's equations are used, which is the reason for their robustness.

%

\section{Field-Source Relations in Lorenz Gauge}

Before we proceed, it is useful to examine the field-source relation of an electromagnetics system.
%
Using the Green's function technique \cite{kong2008ewt,chew1990waves}, the solutions to the above equations are found to be
\begin{align}\label{pezg:eq1b}
    \Phi\prw = \frac{1}{\varepsilon_0} \int_V g(\v r,\v r') \varrho_{\t{imp}}\prpw d\v r' \\
    \v A\prw = \mu_0 \int_V g(\v r,\v r') \v J_{\t{imp}}\prpw d\v r'
    \label{pezg:eq2b}
\end{align}
where $g(\v r,\v r')=\exp(ik_0|\v r-\v r'|)/(4\pi |\v r-\v r'|)$ is the Green's function with $k_0=\omega\sqrt{\mu_0\varepsilon_0}=\omega/c_0$.  It can be shown that the above \eqref{pezg:eq1b} and \eqref{pezg:eq2b} satisfy
\eqref{pezg:eq1a} and  \eqref{pezg:eq2a} by back substitution \cite{kong2008ewt,chew1995waves,ece604lectures}.  It is to be noted that in free-space, the above is the convolution between the Green's function and the source.  Since the Green's function has infinite support, the resultant field also has infinite support, implying that the fields are, in general, nonzero outside the source region.   

\begin{figure}[h!]
    \centering
    \includegraphics[width=1.\linewidth]{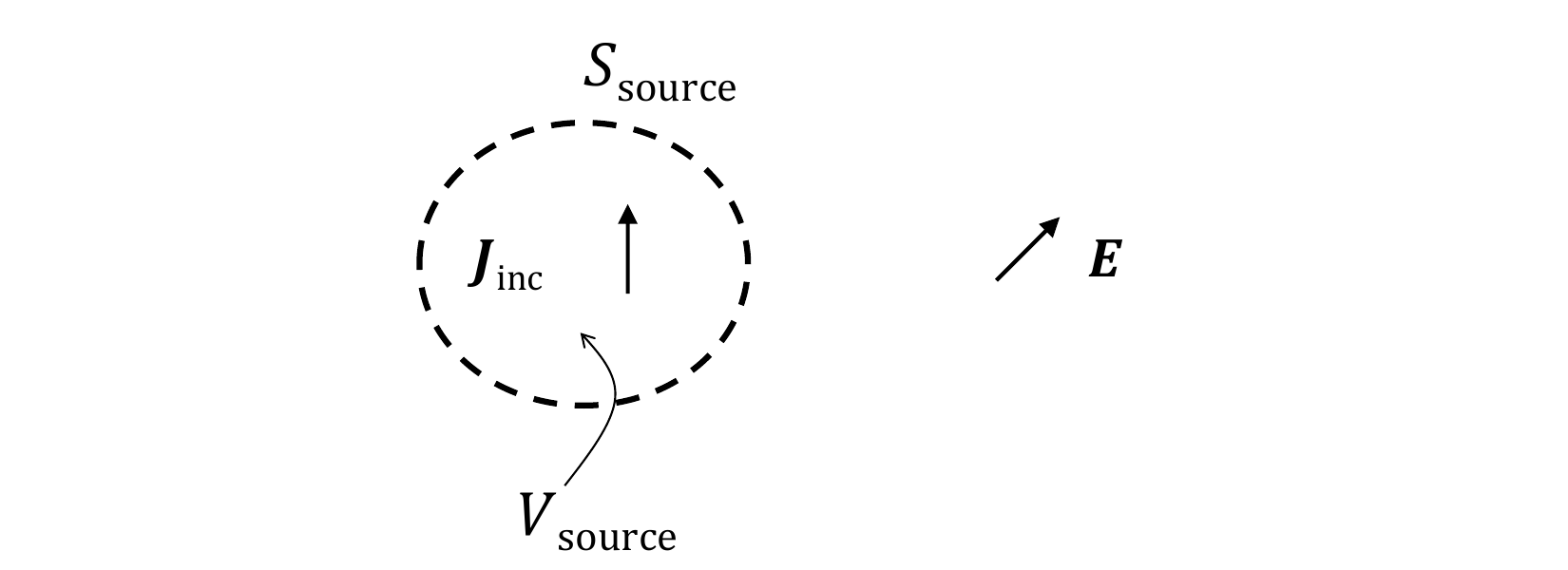}
    \caption{A source bounded with $S_\t{source}$ but radiating to its outside in free space.}\label{figure1}
\end{figure}

Using \eqref{pezg:eq3b}, we get that $\v E\prw=i\omega \v A\prw-\grad \Phi\prw$.  Thus the total electric field $\v E\prw$ is given by
%
\begin{align}\label{pezg:eq3a}
\v E\prw=i\omega \mu_0 \int_V d\v r' g(\v r,\v r')\v J_{\t{imp}}\prpw-\frac{\nabla}{\varepsilon_0}
\int_V d\v r' g(\v r,\v r') \varrho_{\t{imp}}\prpw
\end{align}
At this point, we can give physical interpretation to the above: In the vector potential term or the first term on the \RHS, the current generates the vector potential $\v A$ via \eqref{pezg:eq2b}.  Then the time derivative of the vector potential gives the electric field by induction, which in the frequency domain, $\v E_A=i\omega\v A$.  On the other hand, in the second term, the charge generates the scalar potential $\Phi$ directly, and the negative gradient of it gives the electric field.

In general, by using Helmholtz decomposition, the vector current $\v J_\t{imp}=\v J_{\perp,\t{imp}}
+\v J_{\parallel,\t{imp}}$, where $\div \v J_{\perp,\t{imp}}=0$ and $\curl \v J_{\parallel,\t{imp}}=0$.  Then it can be shown that by Helmholtz decomposition, and letting $\v A=\v A_{\perp}+\v A_\parallel$, where $\div \v A_\perp=0$ and $\curl \v A_\parallel=0$, that
\begin{align}\label{rg:eqn14}
    \v A_\perp \prw = \mu_0 \int_V g(\v r,\v r') \v J_{\perp,{\t{imp}}}\prpw d\v r'\\\label{rg:eqn15}
    \v A_\parallel \prw = \mu_0 \int_V g(\v r,\v r') \v J_{\parallel,{\t{imp}}}\prpw d\v r'
\end{align}
It can be shown that the above equations are divergence-free and curl-free, respectively, by back-substitution.

%

\subsection{Redundancy of the Scalar Potential}

First, we will show that in a homogeneous source-free medium (or outside the source region), the scalar potential is redundant and hence, can be omitted for simplicity.  From the previous equations \eqref{pezg:eq1a} and \eqref{pezg:eq2a}, we have
%
Outside the source region or $V_\t{source}$, ${\bf{J}}_\textrm{inc}=\varrho_\textrm{inc}=0$, and we have $\nabla\cdot{\bf{E}}=0$. Hence, by using \eqref{pezg:eq3} and taking its divergence, we have
\begin{equation}\label{pezg:eq7}
    \nabla\cdot{\bf{E}}=i\omega\nabla\cdot{\bf{A}}-\nabla\cdot\nabla\Phi=0.
\end{equation}
In other words, the non-divergence-free part of $\v A$ is cancelled by the non-divergence-free part of $\grad \Phi$.
To see the physics more clearly, first we decompose $\v A$ into a divergence-free and curl-free parts according to Helmholtz (or Hodge) decomposition \cite{von1902hermann,collin2007foundations} as before or $\v A=\v A_\perp+\v A_\parallel$.
Then \eqref{pezg:eq7} becomes
\begin{equation}\label{pezg:eq8}
\nabla\cdot{\bf E}=i\omega\nabla\cdot{\bf A}_\parallel-\nabla^2\Phi=0
\end{equation}
The above follows because $\div \v A_\perp = 0$.  Also, when $\omega\rightarrow 0$, \eqref{pezg:eq8} goes to zero smoothly without ${\bf A}_\parallel$ or $\Phi$ being divergent.  Or $i\omega \Div\v A_\parallel=\nabla^2 \Phi=-\omega^2\mu_0\varepsilon_0\Phi\rightarrow 0$.
Using the Lorenz gauge given in \eqref{pezg:eq0}, the frequency dependence of the above can be easily verified.
%
%
%
%

In other words, outside the source region, it suffices to have the vector potential $\v A _\perp\ne 0$ outside the source region, or
\begin{align}\label{rg:eqn20}
\v E=i\omega \v A_\perp
\end{align}
The above field is clearly divergence-free corresponding to a field in a source-free region.

\begin{figure}[h!]
    \centering
    \includegraphics[width=1.\linewidth]{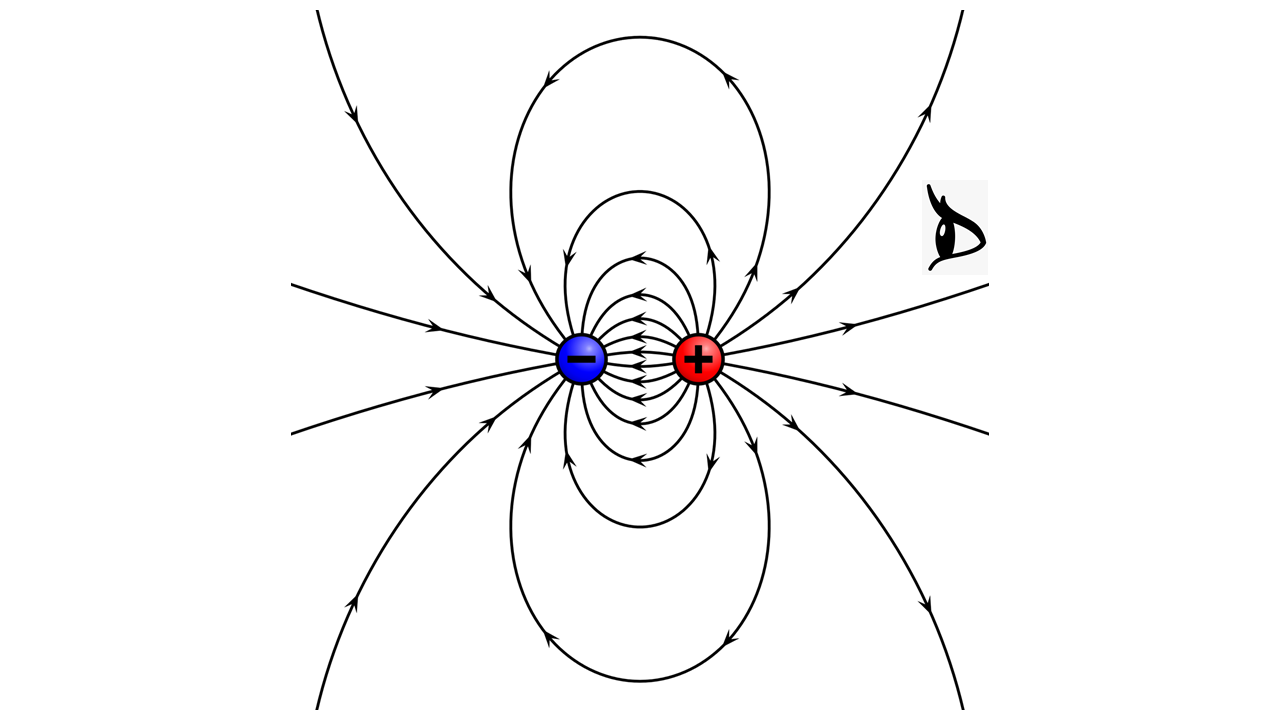}
    \caption{Observing a time-varying dipole field in the near-field regime.  The electric field coming from the scalar potential, $-\grad\Phi$ can be much stronger that the contribution from the vector potential $i\omega \v A$.
    }\label{dipole_observer}
\end{figure}

\section{Physical Interpretation}

It is often said that the radiation gauge is similar to the $\Phi=0$ gauge.  This can be a misunderstanding, as $\Phi\ne 0$ for most problems.  In fact, in the near field, the electric field associate with $\Phi$ can be much larger than that associated with $\v A$ as $\v E=i\omega \v A-\grad \Phi$.  One can surmise the $i\omega \v A$ can be much smaller than $\grad \Phi$ when $\omega$ is small.  This is synonymous with being in the near field, or in the low frequency limit.  

However, in the far-field approximations of both \eqref{pezg:eq1b} and \eqref{pezg:eq2b}, the fields then look like spherical waves or plane waves.  In this case, $i\omega \v A_\parallel$ and $\grad \Phi$ are of the same order, and they almost cancel each other numerically.  Hence, the fields are redundant and we can set $\Phi=\v A_\parallel=0$ with abandon.

\subsection{Homomorphism with Electric Field}

When we let $\v E=i\omega \v A_\perp$, and ignore $\Phi$, the integral equation of scattering using the $\v A$-$\Phi$ formulation is mathematically similar to the electric field integral equation (EFIE) of scattering.  When $\omega\ne 0$, and $\v E$ and $\v A$ are of the same order, and they differ by a phase difference.  Therefore, setting $\Phi=0$ in the $\v A$-$\Phi$ formulation is similar to solving the scattering problem with EFIE.  However, one has to do this with caution when $\omega\rightarrow 0$ since $\v E=i\omega \v A-\grad \Phi$.  When $\Phi=0$, the first term $i\omega \v A$ will have to bear the burden of modeling a static electric field whenever $\omega\rightarrow 0$, since the electric field is non-zero for an electrostatic problem.  This makes $\v A$ diverges as $1/\omega$ when $\omega\rightarrow 0$.

%
%
%

\section{Inhomogeneous Dispersionless Media}


Again, we can expand $\v B$ and $\v E$ as in
\eqref{pezg:eq3} and \eqref{pezg:eq3b} where $\v B=\mu(\v r) \v H$ now.
To satisfy the other two \Meq,
it can be shown that
the equations to be satisfied are \cite{chew2014vector}
\begin{align}\label{rg:eqn22a}
\nabla\cdot\varepsilon(\v r)\nabla\Phi&+\omega^2\varepsilon^2(\v r)\mu(\v r)\Phi=-\varrho_{\t{imp}}
\\\label{rg:eqn23a}
-\nabla\times\mu^{-1}(\v r)\nabla\times{\bf A}&+\varepsilon(\v r)\nabla\left(\frac{1}{\mu(\v r)\varepsilon^2(\v r)}\nabla\cdot\varepsilon(\v r){\bf A}\right)+\omega^2\varepsilon(\v r){\bf A}=-\bf J_{\t{imp}}
\end{align}
In arriving at the above, we have used the generalized Lorenz gauge for inhomogeneous media \cite{chew2014vector}
\begin{align}\label{GLG}
\nabla\cdot\varepsilon(\v r)\mathbf{A}=i\omega\varepsilon^2(\v r)\mu(\v r){\Phi}
\end{align}
Also, it is to be noted that the fields above are implied functions of $\v r$ and $\omega$ as well.

It can be shown that a generalized Helmholtz (or Hodge) decomposition can be used to divide the Hilbert space that $\v A$ lives in, into
 two orthogonal subspaces \cite{zhu2023generalized}. 
   Then the vector potential can be decomposed into \cite{zhu2023generalized}
\begin{align}
\v A=\v A_{\varepsilon\perp}+\v A_\parallel
\end{align}
where $\div \varepsilon\v A_{\varepsilon\perp}=0$ and $\curl \v A_\parallel=0$.  In a word, $\v A_{\varepsilon\perp}$ lives in the null space of the divergence operator, while $\v A_{\parallel}$ lives in the null space of the curl operator.
Moreover, it can be shown \cite{zhu2023generalized} that vectors in these two spaces are $\varepsilon$-orthogonal to each other, or that
\begin{align}
\langle \v A^*_{\varepsilon,\perp},\varepsilon\v A_\parallel\rangle=
\int_V d\v r A^*_{\varepsilon,\perp}\prw \cdot \varepsilon\v A_\parallel\prw= 0
\end{align}
In the above, we have used the reaction inner product that is popular in the electromagnetics literature \cite{rumsey1954reaction,harrington1961time,chew2008integral} where the complex conjugation is explicitly shown.
It is to be noted that this decomposition is independent of frequency since $\varepsilon$ is frequency independent.
%
%
Again, for the inhomogeneous, dispersionless, and source-free media, we can show that the scalar potential is redundant.
In this case, using the generalized Helmholtz decomposition, letting $\v A=\v A_{\varepsilon,\perp} + \v A_\parallel$.  Then requiring $\div\varepsilon\v E=0$, we arrive at
\begin{align}
\div\varepsilon \v E =0&=i\omega \div \varepsilon \v A-\div\varepsilon\grad \Phi\notag\\
&=i\omega \div \varepsilon \v A_\parallel-\div\varepsilon\grad \Phi
\end{align}
By using the generalized Lorenz gauge \eqref{GLG}, and \eqref{rg:eqn22a} outside the source region, it can be shown that the above is zero, satisfying the requirement of a source-free $\v E$ field. Hence, outside the source region, $\Phi$ produces a component of $\v E$ field that together with the contribution of from $\v A_\parallel$ makes $\varepsilon \v E$ divergence free.  To make $\varepsilon \v E$ divergence free outside the source, it suffices to let
\begin{align}\label{rg:eqn33}
\v E=i\omega \v A_{\varepsilon,\perp}
\end{align}
  In other words, both $\Phi$ and $\v A_\parallel$ are redundant outside the source.  They can be set to zero for simplicity, but they are not individually zero.  
  However, since $\Phi$ and $\v A_\parallel$ are produced by finite sources in accordance with \eqref{pezg:eq1b} and \eqref{pezg:eq2b}, just as in the free-space case, we do not expect them to be zero outside their source region.

\section{Conclusion 
}

We have gained a better understanding of the radiation gauge and summarized it in this article.
While this gauge has been called by various names such as Weyl gauge \cite{lancaster2014quantum}, $\Phi=0$ gauge \cite{Gerry2004introductory}, and radiation gauge \cite{jackson1999classical}, there has not been a proper documentation as to when this gauge is valid.  In general, the radiation gauge is safe to use for electrodynamic problems where the scalar potential and vector potential are of equal importance and one is   outside the source region.  

But in the near field or the low frequency regime, one has to use this gauge with caution.  The electrostatic field can be quite strong, and if we let $\v E=i\omega \v A$, because $\v E$ is finite when $\omega\rightarrow 0$,  $\v A\rightarrow 1/\omega$. Thus $\v A$ will become bigger as $\omega\rightarrow 0$ to bear the burden of modeling the electrostatic field.  This is unphysical because it diverges.   Similar statement can be made when the medium is inhomogeneous and dispersionless.

\bibliography{embibr_rg}

\begin{thebibliography}{10}
\providecommand{\url}[1]{#1}
\csname url@samestyle\endcsname
\providecommand{\newblock}{\relax}
\providecommand{\bibinfo}[2]{#2}
\providecommand{\BIBentrySTDinterwordspacing}{\spaceskip=0pt\relax}
\providecommand{\BIBentryALTinterwordstretchfactor}{4}
\providecommand{\BIBentryALTinterwordspacing}{\spaceskip=\fontdimen2\font plus
\BIBentryALTinterwordstretchfactor\fontdimen3\font minus \fontdimen4\font\relax}
\providecommand{\BIBforeignlanguage}[2]{{%
\expandafter\ifx\csname l@#1\endcsname\relax
\typeout{** WARNING: IEEEtran.bst: No hyphenation pattern has been}%
\typeout{** loaded for the language `#1'. Using the pattern for}%
\typeout{** the default language instead.}%
\else
\language=\csname l@#1\endcsname
\fi
#2}}
\providecommand{\BIBdecl}{\relax}
\BIBdecl

\bibitem{jackson1999classical}
J.~D. Jackson, \emph{Classical electrodynamics}.\hskip 1em plus 0.5em minus 0.4em\relax John Wiley \& Sons, 1999.

\bibitem{yaghjian2014reflections}
A.~D. Yaghjian, ``Reflections on {M}axwell's treatise,'' \emph{Progress In Electromagnetics Research}, vol. 149, pp. 217--249, 2014.

\bibitem{chew2014vector}
W.~C. Chew, ``Vector potential electromagnetics with generalized gauge for inhomogeneous media: {F}ormulation,'' \emph{Progress In Electromagnetics Research}, vol. 149, pp. 69--84, 2014.

\bibitem{li2009finite}
Y.~Li, S.~Sun, and W.~C. Chew, ``Finite element implementation of the generalized-{L}orenz gauged,'' \emph{IEEE Trans. Antennas Propag}, vol.~57, pp. 3594--3601, 2009.

\bibitem{li2015vectorial}
Y.-L. Li, S.~Sun, Q.~I. Dai, and W.~C. Chew, ``Vectorial solution to double curl equation with generalized {Coulomb} gauge for magnetostatic problems,'' \emph{IEEE Transactions on Magnetics}, vol.~51, no.~8, pp. 1--6, 2015.

\bibitem{li2016finite}
------, ``Finite element implementation of the generalized-{L}orenz gauged {A-Phi} formulation for low-frequency circuit modeling,'' \emph{IEEE Transactions on Antennas and Propagation}, vol.~64, no.~10, pp. 4355--4364, 2016.

\bibitem{li2016loop}
Y.-L. Li, S.~Sun, and Z.-H. Ma, ``Loop-based flux formulation for three-dimensional magnetostatic problems,'' \emph{Applied Computational Electromagnetics Society Journal}, vol.~31, no.~11, 2016.

\bibitem{liu_qin2018potential}
Q.~S. Liu, S.~Sun, and W.~C. Chew, ``A potential-based integral equation method for low-frequency electromagnetic problems,'' \emph{IEEE Transactions on Antennas and Propagation}, vol.~66, no.~3, pp. 1413--1426, 2018.

\bibitem{roth2020radiation}
T.~E. Roth and W.~C. Chew, ``Radiation gauge potential-based time domain integral equations for penetrable regions,'' \emph{Progress In Electromagnetics Research}, vol. 168, pp. 73--86, 2020.

\bibitem{CHEW2016quantum}
W.~C. {Chew}, A.~Y. {Liu}, C.~{Salazar-Lazaro}, and W.~E.~I. {Sha}, ``Quantum electromagnetics: A new look---{P}art {I} and {P}art {II},'' \emph{J. Multiscale and Multiphys. Comput. Techn.}, vol.~1, pp. 73--97, 2016.

\bibitem{lancaster2014quantum}
T.~Lancaster and S.~J. Blundell, \emph{Quantum field theory for the gifted amateur}.\hskip 1em plus 0.5em minus 0.4em\relax OUP Oxford, 2014.

\bibitem{vico2016decoupled}
F.~Vico, M.~Ferrando, L.~Greengard, and Z.~Gimbutas, ``The decoupled potential integral equation for time-harmonic electromagnetic scattering,'' \emph{Communications on Pure and Applied Mathematics}, vol.~69, no.~4, pp. 771--812, 2016.

\bibitem{liu2018potential}
Q.~S. Liu, S.~Sun, and W.~C. Chew, ``A potential-based integral equation method for low-frequency electromagnetic problems,'' \emph{IEEE Transactions on Antennas and Propagation}, vol.~66, no.~3, pp. 1413--1426, 2018.

\bibitem{roth2018development}
T.~E. Roth and W.~C. Chew, ``Development of stable {A-$\Phi$} time-domain integral equations for multiscale electromagnetics,'' \emph{IEEE Journal on Multiscale and Multiphysics Computational Techniques}, vol.~3, pp. 255--265, 2018.

\bibitem{roth2020stability}
------, ``Stability analysis and discretization of {A-$\Phi$} time domain integral equations for multiscale electromagnetics,'' \emph{Journal of Computational Physics}, vol. 408, p. 109102, 2020.

\bibitem{boyd2011distributed}
S.~Boyd, N.~Parikh, E.~Chu, B.~Peleato, J.~Eckstein \emph{et~al.}, ``Distributed optimization and statistical learning via the alternating direction method of multipliers,'' \emph{Foundations and Trends{\textregistered} in Machine learning}, vol.~3, no.~1, pp. 1--122, 2011.

\bibitem{kong2008ewt}
J.~A. Kong, \emph{Electromagnetic Wave Theory}.\hskip 1em plus 0.5em minus 0.4em\relax EMW Publishing, 2008, also 1985.

\bibitem{ece604lectures}
W.~C. Chew, ``{Electromagnetic Field Theory, {Purdue U}},'' \url{ https://engineering.purdue.edu/wcchew/ece604f23/ }.

\bibitem{hiptmair1998multigrid}
R.~Hiptmair, ``Multigrid method for {M}axwell's equations,'' \emph{SIAM Journal on Numerical Analysis}, vol.~36, no.~1, pp. 204--225, 1998.

\bibitem{axelson1984finite}
O.~Axelson and V.~Barker, \emph{Finite element solution of boundary value problems}.\hskip 1em plus 0.5em minus 0.4em\relax Academic Press, New York, 1984.

\bibitem{chew1990waves}
W.~C. Chew, \emph{Waves and Fields in Inhomogeneous Media, 378$\pm$381}.\hskip 1em plus 0.5em minus 0.4em\relax Van Nostrand, 1990.

\bibitem{chew1995waves}
------, \emph{Waves and fields in inhomogeneous media}.\hskip 1em plus 0.5em minus 0.4em\relax IEEE Press, 1995, also 1990.

\bibitem{von1902hermann}
H.~L. von Helmholtz and L.~Koenigsberger, \emph{Hermann von {H}elmholtz}.\hskip 1em plus 0.5em minus 0.4em\relax F. Vieweg, 1902.

\bibitem{collin2007foundations}
R.~E. Collin, \emph{Foundations for microwave engineering}.\hskip 1em plus 0.5em minus 0.4em\relax John Wiley \& Sons, 2007, also 1966.

\bibitem{zhu2023generalized}
J.~Zhu, T.~E. Roth, D.-Y. Na, and W.~C. Chew, ``Generalized {H}elmholtz decomposition for modal analysis of electromagnetic problems in inhomogeneous media,'' \emph{IEEE Journal on Multiscale and Multiphysics Computational Techniques}, 2023.

\bibitem{rumsey1954reaction}
V.~Rumsey, ``Reaction concept in electromagnetic theory,'' \emph{Physical Review}, vol.~94, no.~6, p. 1483, 1954.

\bibitem{harrington1961time}
R.~F. Harrington, \emph{Time-harmonic electromagnetic fields}.\hskip 1em plus 0.5em minus 0.4em\relax McGraw-Hill, 1961.

\bibitem{chew2008integral}
W.~C. Chew, M.~S. Tong, and B.~Hu, ``Integral equation methods for electromagnetic and elastic waves,'' \emph{Synthesis Lectures on Computational Electromagnetics}, vol.~3, no.~1, pp. 1--241, 2008.

\bibitem{Gerry2004introductory}
C.~Gerry and P.~Knight, \emph{Introductory Quantum Optics}.\hskip 1em plus 0.5em minus 0.4em\relax Cambridge, UK: Cambridge University Press, 2004.

\end{thebibliography}

\bibliographystyle{IEEEtran}

\end{document}